\documentclass{pi2005}
\usepackage{amsmath}
\usepackage{amsfonts}
\usepackage{amssymb}
\usepackage{graphicx}
\usepackage{cite}

\def\tr{\operatorname{tr}}

\slacs{.4ex}

\begin{document}
\title{Path integral evaluation of the kinetic isotope effects based on the quantum instanton approximation}
\authori{Ji\v{r}\'{\i} Van\'{\i}\v{c}ek\,\footnote{Electronic mail: vanicek@post.harvard.edu}
and William H. Miller}
\addressi{Department of Chemistry and Kenneth S. Pitzer Center for Theoretical Chemistry,\\
University of California, Berkeley, CA 94720, USA}
\authorii{}    \addressii{}
%
\headauthor{Ji\v{r}\'{\i} Van\'{\i}\v{c}ek and William H. Miller}
\headtitle{Quantum instanton evaluation of the kinetic isotope
effects}
\lastevenhead{Ji\v{r}\'{\i} Van\'{\i}\v{c}ek et al.: Quantum instanton evaluation of the kinetic isotope effects}
\maketitle

\begin{abstract}
A general method for computing kinetic isotope effects is
described. The method uses the quantum-instanton approximation and
is based on the thermodynamic integration with respect to the mass
of the isotopes and on the path-integral Monte-Carlo evaluation of
relevant thermodynamic quantities. The central ingredients of the
method are the Monte-Carlo estimators for the logarithmic
derivatives of the partition function and the delta--delta
correlation function. Several alternative estimators for these
quantities are described here and their merits are compared on the
benchmark hydrogen-exchange reaction,
H+H$_{2}\rightarrow$H$_{2}$+H on the Truhlar--Kuppermann potential
energy surface. Finally, a qualitative discussion of issues
arising in many-dimensional systems is provided.
\end{abstract}

\section{Introduction}

Measurement and theoretical predictions of kinetic isotope effects
belong among the main tools of chemical kinetics. Kinetic isotope
effect (KIE) is defined as the ratio $k_{A}/k_{B}$ of rate
constants for two isotopomers $A$ and $B$. Isotopomers $A$ and $B$
are two chemical species differing only by replacing a group of
atoms in chemical species $A$ by their isotopes in species $B$.
Recently, observation of anomalously large KIEs has helped prove
importance of quantum effects in enzymatic reactions at
physiological (i.e. surprisingly high) temperatures
\cite{kohen:1999}. This and similar results have changed our
understanding of enzymatic catalysis and spurred an active
experimental and theoretical research in the last several years.

Since the early days of chemical kinetics, KIEs have been
predominantly described from the perspective of the
transition-state theory (TST) \cite{melander:1960,collins:1971}.
This theory is intrinsically classical, although various quantum
\textquotedblleft corrections\textquotedblright\ have been
incorporated in it over time. These include corrections due to the
zero-point-energy effects, high-temperature Wigner tunneling
correction \cite{melander:1960,collins:1971}, and various
semiclassical approximations for treating the tunneling at low
temperatures \cite{miller:1975,marcus:1977,truhlar:1984}. On the
other end of the spectrum are exact quantum-mechanical methods for
computing rate constants and KIEs \cite{miller:1998}, but in
general these are not feasible for systems with many degrees of
freedom. One therefore resorts to various approximations that make
a computation practicable but are less severe than the TST. Among
these belongs a variety of quantum transition-state theories
\cite{voth:1989,hansen:1996,truong:1993}, the most recent of which
is the quantum instanton (QI)\ approximation \cite{miller:2003},
motivated by an earlier semiclassical instanton model
\cite{miller:1975}. In this contribution, we describe a method
\cite{vanicek:2005}, based on the QI approximation, for computing
KIEs directly, rather than via computing the rate constants for
the two isotopomers first. Because of the ultimate goal of
applying a similar method to enzymatic reactions, the method is
implemented using a general path-integral approach that scales
favorably with the number of degrees of freedom. Several
alternative estimators for relevant quantities have been developed
\cite{vanicek:2005,vanicek:2005b} and their relative merits are
compared in this contribution on the benchmark hydrogen
atom--diatom exchange reaction.

\section{Quantum instanton approximation for the kinetic isotope effects}

The quantum-instanton approximation for the rate constant was
introduced in Ref. \cite{miller:2003}. A simpler alternative
derivation \cite{ceotto:2003} described in detail in Ref.
\cite{vanicek:2005}, starts with the Miller--Schwartz--Tromp
formula \cite{miller:1983} for the thermal rate constant~$k$,
\begin{equation}
k\,Q_r=\int_0^{\infty}\D t\,C_{\mathrm{ff}}(t)\,.
\label{miller_schwartz_tromp}
\end{equation}
Here $Q_r$ is the reactant partition function (per unit volume for
bimolecular reactions) and $C_{\mathrm{ff}}(t)$ is the symmetrized
flux--flux correlation function,
\begin{equation}
C_{\mathrm{ff}}(t)=\tr\left(\E^{-\beta\hat{H}/2}\hat{F}_{a}\E^{-\beta\hat{H}/2}
\E^{\I\hat{H}t/\hbar}\hat{F}_{b}\E^{-\I\hat{H}t/\hbar}\right)
\label{cff}
\end{equation}
with Hamiltonian operator $\hat{H}$ and flux operator
$\hat{F}_{\gamma}$ denoting the flux through the dividing surface
$\gamma=a,b$. Quantum instanton expression follows by multiplying
and dividing the integrand of Eq.(\ref{miller_schwartz_tromp}) by
the ``delta--delta" correlation function $C_{\mathrm{dd}}(t)$
defined below in Eq. (\ref{cdd}), assuming that
$C_{\mathrm{ff}}(t)/C_{\mathrm{dd}}(t)$ varies slowly compared
with $C_{\mathrm{dd}}(t)$, and applying the steepest descent
approximation to the resulting integral. Assuming further that the
stationary-phase point is at $t=0$, we obtain the QI thermal rate
constant,
\begin{equation}
k_{\text{QI}}=\frac{1}{Q_{r}}\,C_{\mathrm{ff}}(0)\,\frac{\sqrt{\pi}}2\,
\frac{\hbar}{\Delta H}\,.
\label{k_QI}
\end{equation}
Here $\Delta H$ is a specific type of energy variance
\cite{yamamoto:2004},
\begin{equation}
\Delta
H=\hbar\left[\frac{-\ddot{C}_{\mathrm{dd}}(0)}{2C_{\mathrm{dd}}(0)}\right]^{1/2},
\label{dh}
\end{equation}
and the delta--delta correlation function $C_{\mathrm{dd}}(t)$ is
defined \cite{vanicek:2005,yamamoto:2004} as
\begin{equation}
C_{\mathrm{dd}}(t)=\tr\left(\E^{-\beta\hat{H}/2}\hat{\Delta}_{a}\E^{-\beta\hat{H}/2}
\E^{\I\hat{H}t/\hbar}\hat{\Delta}_{b}\E^{-\I\hat{H}t/\hbar}\right).
\label{cdd}
\end{equation}
The generalized delta operator $\hat{\Delta}$ will be defined
below in Eq.~(\ref{gen_delta}).

In applying the QI approximation to the KIEs, it is useful to
consider a continuous change of the isotope mass. If the two
isotopomers are $A$ and $B$, a real parameter $\lambda\in[0,1]$
can be defined such that
$$
m_i(\lambda)=m_{A,i}(1-\lambda)+m_{B,i}\lambda\,,
$$
where $m_{A,i}$ and $m_{B,i}$ are the masses of the $i$th atom in
the isotopomers $A$ and $B$, respectively. Within the QI
approximation (\ref{k_QI}), the KIE can be expressed as
\begin{equation}
\text{KIE}_{\text{QI}}=\frac{k_{\text{QI}}(0)}{k_{\text{QI}}(1)}=
\frac{Q_{r}(1)}{Q_{r}(0)}\times\frac{\Delta H(1)}{\Delta
H(0)}\times \frac{C_{\mathrm{dd}}(0)}{C_{\mathrm{dd}}(1)}\times
\frac{C_{\mathrm{ff}}(0)/C_{\mathrm{dd}}(0)}
{C_{\mathrm{ff}}(1)/C_{\mathrm{dd}}(1)}\,,
\label{qi_kie_1}
\end{equation}
where the argument denotes the value of $\lambda$ and for
simplicity the time argument of the correlation functions has been
omitted since it is always $t=0$ in the QI approximation. Also,
for convenience, both numerator and denominator have been divided
by $C_{\mathrm{dd}}(\lambda)$.

Four types of quantities must be evaluated in order to compute the
KIEs from Eq. (\ref{qi_kie_1}): the ratio of the partition
functions $Q_r(1)/Q_r(0)$, the ratio of the delta correlation
functions $C_{\mathrm{dd}}(1)/C_{\mathrm{dd}}(0)$, and the energy
variance $\Delta H(\lambda)$ and the ``velocity" factor
$C_{\mathrm{ff}}(\lambda)/C_{\mathrm{dd}}(\lambda)$ for
$\lambda=0$ and 1. The last two quantities are in the form of
thermodynamic averages (for a given $\lambda$) and therefore can
be directly computed by Metropolis Monte-Carlo techniques; the
relevant estimators have been derived in Refs.
\cite{yamamoto:2004,zhao:2004}. The most general forms are listed
in Ref. \cite{vanicek:2005}. The first two quantities cannot be
evaluated directly since they are ratios of quantities for two
different values of $\lambda$.

An elegant solution exists, however. Here is where considering a
continuous isotope change (using a parameter $\lambda$) becomes
useful: instead of computing the ratios directly, we use the
\textit{thermodynamic integration} idea \cite{chandler:1987},
applied to the parameter $\lambda$ (i.e., to the masses of the
isotopes instead to the usual inverse temperature $\beta$). We can
express the two ratios as an exponential of the integrals of
logarithmic derivatives,
\begin{eqnarray}
\frac{Q_r(1)}{Q_r(0)}&=&\exp\left[\int_0^1\D\lambda\,\frac{\D\log
Q_r(\lambda)}{\D\lambda}\right],\label{charging_Qr}\\
\frac{C_{\mathrm{dd}}(1)}{C_{\mathrm{dd}}(0)}&=&
\exp\left[\int_0^1\D\lambda\,\frac{\D\log
C_{\mathrm{dd}}(\lambda)}{\D\lambda}\right]. \label{charging_Cdd}
\end{eqnarray}
Since the logarithmic derivatives can be expanded as
$$
\frac{\D\log\rho(\lambda)}{\D\lambda}=\frac{\D\rho(\lambda)/\D\lambda}{\rho(\lambda)}\,,
$$
they are normalized quantities (thermodynamic averages) which can
be directly computed by the Metropolis algorithm. We can compute
ratios of both the reactant partition functions and the
delta--delta correlation functions at $\lambda=0$ and 1, by
computing the values of the corresponding logarithmic derivatives
for enough values $\lambda$ between 0 and 1, and then by
integrating over $\lambda$ and exponentiating, according to Eqs.
(\ref{charging_Qr}) and (\ref{charging_Cdd}).

In fact, in a cruder version of the QI method, called the simplest
quantum instanton (SQI) approximation \cite{miller:2003}, the
ratios of the partition and delta--delta correlation functions are
all we need, since within that approximation, the kinetic isotope
effect is just
\begin{equation}
\text{KIE}_{\text{SQI}}=\frac{Q_{r}(1)}{Q_{r}(0)}\times
\frac{C_{\mathrm{dd}}(0)}{C_{\mathrm{dd}}(1)}\,.
\label{qi_kie_2}
\end{equation}

The relevant estimators for the logarithmic derivatives have been
derived in Refs. \cite{vanicek:2005,vanicek:2005b}. In Ref.
\cite{vanicek:2005}, \textit{thermodynamic} estimators have been
derived that differentiate the kinetic part of the action; in Ref.
\cite{vanicek:2005b}, \textit{virial} estimators have been derived
that differentiate the potential part of the action. In both
cases, the derivations have been done for general systems with $N$
atoms in $d$ dimensions, even for cases with unbound degrees of
freedom (such as the center-of-mass motion). In the next section,
we present a simplified derivation of these estimators for a
single particle in a one-dimensional external potential. This
choice significantly simplifies notation, but preserves the main
ingredients of the many-dimensional derivation.

\section{Estimators for the logarithmic derivatives of \bmth{Q_r} and \bmth{C_{\mathrm{dd}}}}

Below we derive three types of estimators for the logarithmic
derivatives of both $Q_r$ and $C_{\mathrm{dd}}$. We refer to the
three types of estimators as thermodynamic, virial, and
generalized virial because of their resemblance to corresponding
thermodynamic \cite{barker:1979}, virial \cite{herman:1982}, and
generalized virial \cite{parrinello:1984} estimators for the
kinetic energy.

\subsection{Partition function}

Let us consider a single particle of mass $m$ in a one-dimensional
potential $V(r)$. Since we have only one mass $m$, we do not need
to define an additional parameter $\lambda$: we can just take $m$
itself to be the parameter for the thermodynamic integration. The
PI representation of the partition function is
\begin{eqnarray}
Q_r&\simeq&\left(\frac{Pm}{2\pi\hbar^2\beta}\right)^{P/2} \int\D
r^{(1)}\cdots\int\D r^{(P)}\rho_r\left(\{r^{(s)}\}\right),
\label{qr_pi_cartesian}\\
\rho_{r}\left(\{r^{(s)}\}\right)&=&\exp\left[-\beta\Phi(\{r^{(s)}\})\right],\\
\Phi&=&\frac{Pm}{2\pi\hbar^2\beta^2}\sum_{s=1}^P\left(r^{(s)}-r^{(s-1)}\right)^2+
\frac{1}{P}\sum_{s=1}^{P}V\left(r^{(s)}\right).\nonumber
\end{eqnarray}
Here $s=1,\ldots,P$, denotes the beads of the discretized paths
($s=0$ is identical to $s=P$). In general, we will obtain the
estimators for the logarithmic derivative directly, by computing
the logarithmic derivative $\dfrac{\D\log Q_r}{\D m}=
\dfrac1{Q_r}\,\dfrac{\D Q_r}{\D m)}$ of the particular form of the
discretized PI. Applying this approach to the PI
(\ref{qr_pi_cartesian}), we obtain the \textit{thermodynamic}
estimator
\begin{eqnarray}
\frac{\D\log Q_r}{\D m}&=&
\frac{P}{2m}-\beta\left\langle\frac{\partial\Phi}{\partial m}\right\rangle_{\rho_{r}},
\label{qr_est_thermodynamic}\\
\frac{\partial\Phi}{\partial m}&=&
\frac{P}{2\pi\hbar^2\beta^2}\sum_{s=1}^{P}\left(r^{(s)}-r^{(s-1)}\right)^2.\nonumber
\end{eqnarray}

Above, $\left\langle A(\left\{r^{(s)}\right\}\right\rangle_{\rho}$
denotes the average over paths weighted with the weight $\rho$,
$$
\left<A\left(\left\{r^{(s)}\right\}\right)\right>_{\rho}\equiv
\frac{\int\D r^{(1)}\cdots\int\D r^{(P)}
A\left(\left\{r^{(s)}\right\}\right)\rho\left(\left\{r^{(s)}\right\}\right)}
{\int\D r^{(1)}\cdots\int\D r^{(P)}
\rho\left(\left\{r^{(s)}\right\}\right)}\,.
$$

Alternatively, we can define new, mass-scaled coordinates as
\begin{equation}
x\equiv m^{1/2}r\,.
\label{rescaling}
\end{equation}
In these new coordinates, the partition function becomes
\begin{eqnarray}
Q_r&\simeq&\left(\frac{P}{2\pi\hbar^2\beta}\right)^{P/2}
\int\D x^{(1)}\cdots\int\D x^{(P)}\E^{-\beta\Phi}\,,\label{qr_pi_rescaled}\\
\Phi&=&\frac{P}{2\pi\hbar^2\beta^2}\sum_{s=1}^P\left(x^{(s)}-x^{(s-1)}\right)^2+
\frac1P\sum_{s=1}^{P}V\left(m^{-1/2}x^{(s)}\right).\nonumber
\end{eqnarray}
Simplest \textit{virial} estimator for the logarithmic derivative
can again be derived by direct differentiation of PI
(\ref{qr_pi_rescaled}),
\begin{eqnarray}
\frac{\D\log Q_r}{\D m}&=&-\frac{\beta}P\left<\sum_{s=1}^P
\frac{\partial V\left[m^{-1/2}x^{(s)}\right]} {\partial m}\right>_{\rho_r}=\label{qr_est_virial}\\
&=&-\frac{\beta}P\left<\sum_{s=1}^P\frac{\partial V\left[(m+\Delta
m)^{-1/2}m^{1/2}r^{(s)}\right]} {\partial\Delta m}\biggl|_{\Delta m=0}\right> _{\rho_{r}}=\nonumber\\
&=&\frac{\beta}{2P}\left<\sum_{s=1}^{P}r^{(s)}\frac{\partial
V\left(r^{(s)}\right)}{\partial
r^{(s)}}\right>_{\rho_{r}}.\nonumber
\end{eqnarray}
Above are three estimators for the logarithmic derivative: the
first one suitable if the MC simulation is done in mass-scaled
coordinates $x$, the other two for original Cartesian coordinates
$r$. The first two suggest evaluation of the derivative
numerically, by finite differences, which will be in fact, more
efficient in many dimensional systems than the analytical third
expression that requires the knowledge of the gradient of the
potential. Only for systems with few degrees of freedom and
available gradient of the potential, the third expression may be
preferable. The trick of using numerical derivatives with respect
to a single parameter was originally used by Predescu for
computing heat capacities \cite{predescu:2003} and higher temporal
derivatives of the flux--flux correlation function
\cite{predescu:2004} where the parameters were the inverse
temperature and the imaginary time, respectively.

The simplest virial estimators (\ref{qr_est_virial}) have one
shortcoming compared to the thermodynamic estimators, namely, they
only work in bound systems. This can be immediately seen by
considering a free particle with $V(r)=0$. This shortcoming can be
remedied if the rescaling is done only after subtracting an
arbitrarily chosen (but fixed) slice from the remaining $P-1$
slices. To be more explicit, let us define relative coordinates as
$$
y^{(s)}\equiv
r^{(s)}-r^{(P)}\quad\text{for~}\;s=1,\,\ldots,\,P-1\,.
$$
Since the Jacobian of the transformation is unity, we have
\begin{eqnarray}
Q_r&=&\left(\frac{Pm}{2\pi\hbar^2\beta}\right)^{P/2}\int\D y^{(1)}
\cdots\int\D y^{(P-1)}\int\D r^{(P)}\E^{-\beta\Phi}\,,\label{qr_pi_relative}\\
\Phi&=&\frac{Pm}{2\pi\hbar^2\beta^2}\biggl[\left(y^{(1)}\right)^2+
\sum_{s=2}^{P-1}\left(y^{(s)}-y^{(s-1)}\right)^2+
\left(y^{(P-1)}\right)^2\biggr]+\nonumber\\
&&+\frac1P\biggl[\,\sum_{s=1}^{P-1}V(r^{(P)}+y^{(s)})+V(r^{(P)})\biggr].\nonumber
\end{eqnarray}
Now we define mass-scaled coordinates as
\begin{equation}
x^{(s)}\equiv
m^{1/2}y^{(s)}=m^{1/2}(r^{(s)}-r^{(P)})\quad\text{for~}\;s=1,\,\ldots,\,P-1\,.
\label{rescaling_subtract}
\end{equation}
In these coordinates, the partition function becomes
\begin{eqnarray}
Q_r&=&\left(\frac{P}{2\pi\hbar^2\beta}\right)^{P/2}m^{1/2} \int\D
x^{(1)}\cdots\int\D x^{(P-1)}\int\D r^{(P)} \E^{-\beta\Phi}\,,\label{qr_pi_relative_rescaled}\\
\Phi&=&\frac{P}{2\pi\hbar^2\beta^2}\biggl[\left(x^{(1)}\right)^2+
\sum_{s=2}^{P-1}\left(x^{(s)}-x^{(s-1)}\right)^2+
\left(x^{(P-1)}\right)^2\biggr]+\nonumber\\
&&+\frac1P\biggl[\,\sum_{s=1}^{P-1}V(r^{(P)}+m^{-1/2}x^{(s)})+V(r^{(P)})\biggr].\nonumber
\end{eqnarray}
The \textit{generalized virial} estimator for the logarithmic
derivative follows by differentiating the PI expression
(\ref{qr_pi_relative_rescaled}),
\begin{equation}
\label{qr_est_virial_generalized}
\begin{array}{l}
\disty \frac{\D\log Q_r}{\D m}=
\frac{1}{2m}-\frac{\beta}{P}\biggl<\sum _{s=1}^P \frac{\partial
V\left[r^{(P)}+m^{-1/2}x^{(s)}\right]}
{\partial m}\biggr>_{\rho_{r}}=\\[12pt]
\disty\qquad =\frac{1}{2m}-\frac{\beta}{P}\biggl<\sum_{s=1}^P
\frac{\partial V\left[r^{(P)}+(m+\Delta m)^{-1/2}m^{1/2}
\left(r^{(s)}-r^{(P)}\right)\right]}{\partial\Delta
m}\biggl|_{\Delta m=0}\biggr>_{\rho_r}=\\[12pt]
\disty\qquad =\frac1{2m}+\frac{\beta}{2P}\biggl<
\sum_{s=1}^{P}\frac{\partial V\left(r^{(s)}\right)} {\partial
r^{(s)}}(r^{(s)}-r^{(P)})\biggr>_{\rho_r}\,.
\end{array}
\end{equation}
Since we have chosen the slice $s=P$ arbitrarily, we can do the
same for any slice $s$, derive a corresponding estimator, and then
take an average of these estimators. The result is
\begin{eqnarray}
\frac{\D\log Q}{\D m}&=&\frac{1}{2m}+\frac{\beta}{2P}\biggl<
\sum_{s=1}^P\frac{\partial V\left(r^{(s)}\right)} {\partial
r^{(s)}}\,(r^{(s)}-r^{c})\biggr>_{\rho_r}\,,\label{qr_est_virial_centroid}\\
r^c&\equiv&\frac{1}{P}\sum_{s=1}^{P}r^{(s)}\,.\nonumber
\end{eqnarray}
and in general, we can replace $r^{(P)}$ in all three forms
(\ref{qr_est_virial_generalized}) of the estimator by $r^c$.

Since the number of slices $P$ appears explicitly only in the
denominator of the generalized virial estimator
(\ref{qr_est_virial_generalized}) or
(\ref{qr_est_virial_centroid}), the statistical error should be
independent of $P$ for a fixed number of Monte-Carlo samples.  On
the other hand, $P$ appears explicitly in the numerator of the
thermodynamic estimator (\ref{qr_est_thermodynamic}), so the error
is expected to grow with $P$. This will be confirmed in the
numerical example in Section 4.

\subsection{Delta\bmth{-}delta correlation function}

The derivation for $C_{\mathrm{dd}}$ is similar. However, due to
the constraints to the two dividing surfaces, a new term appears
in the estimator. The PI representation of $C_{\mathrm{dd}}$ is
\begin{eqnarray}
C_{\mathrm{dd}}&\simeq&\left(\frac{Pm}{2\pi\hbar^2\beta}\right)^{P/2}
\int\D r^{(1)}\cdots\int\D r^{(P)}
\rho^{\ddag}\left(\{r^{(s)}\}\right),\label{cdd_pi_cartesian}\\
\rho^{\ddag}\left(\{r^{(s)}\}\right)&=&
\Delta\left[\xi_a\left(r^{(0)}\right)\right]
\Delta\left[\xi_b\left(r^{(P/2)}\right)\right]\E^{-\beta\Phi}.\nonumber
\end{eqnarray}
The generalized delta function $\Delta$ is defined \cite{yamamoto:2004} as
\begin{equation}
\Delta\bigl[\xi(r)\bigr]\equiv\left\vert\frac{m}{\partial_r\xi}\right\vert^{1/2}
\delta\bigl[\xi(r)\bigr]\,.
\label{gen_delta}
\end{equation}
For numerical purposes, it is convenient to replace the strict
delta function by a Gaussian approximation \cite{yamamoto:2004},
\begin{eqnarray}
\Delta\left[\xi\left(r^{s)}\right)\right]&\approx&
\tilde{\Delta}\left[\xi\left(  \bar{r}^{(s)}\right)\right],\label{gauss_delta}\\
\bar{r}^{(s)}&\equiv&\frac{1}{2}\,\left(r^{(s)}+r^{(s+1)}\right),\nonumber\\
\tilde{\Delta}\bigl[\xi(r)\bigr]&\equiv&
\left(\frac{2P}{\pi\hbar^2\beta}\right)^{1/2}
\exp\left\{-\frac{2Pm}{\hbar^2\beta}\left[
frac{\xi(r)}{\partial_r\xi(r)}\right]^2\right\}.\nonumber
\end{eqnarray}
We can define an effective action
$\Phi_{\text{eff}}=\Phi+V_{\text{constr}}$, which includes the
constraint potential
\begin{equation}
V_{\text{constr}}=\frac{2Pm}{\hbar^2\beta^2}\left[\frac{\xi(r)}
{\partial_r\xi(r)}\right]^2.
\label{v_constr}
\end{equation}
The logarithmic derivative will have one extra term due to this
constraint potential,
\begin{equation}
\frac{\D\log C_{\mathrm{dd}}}{\D m}=\frac{\D\log Q_r}{\D m}
\left(\rho_r\rightarrow\rho^{\ddag}\right)-
\beta\left<\frac{\partial V_{\text{constr}}} {\partial
m}\right>_{\rho^{\ddag}}\,,
\label{logderivative_cdd}
\end{equation}
where $\dfrac{\D\log Q_r}{\D
m}\left(\rho_r\rightarrow\rho^{\ddag}\right)$ denotes that the
corresponding estimator for $Q_r$ given in
Eq.~(\ref{qr_est_thermodynamic}), (\ref{qr_est_virial}), or
(\ref{qr_est_virial_generalized}) should be used except that the
sampling is done according to weight $\rho^{\ddag}$ instead of
$\rho_r$. Using the PI representation (\ref{cdd_pi_cartesian}) of
$C_{\mathrm{dd}}$ in Cartesian coordinates, we find the additional
term from Eq.~(\ref{logderivative_cdd}) to the thermodynamic
estimator (\ref{qr_est_thermodynamic}) to be
\begin{equation}
\frac{\partial V_{\text{constr}}}{\partial m}=
\frac{2P}{\hbar^2\beta^2}\left[\frac{\xi(r)}{\partial_r\xi(r)}\right]^2.
\label{cdd_est_thermodynamic}
\end{equation}
Rescaling coordinates according to Eq. (\ref{rescaling}) gives a
constraint potential
$$
V_{\text{constr}}=\frac{2Pm}{\hbar^2\beta^2}\left[\frac{\xi\left(m^{-1/2}x\right)}
{\partial_r\xi(m^{-1/2}x)}\right]^2\,.
$$
The additional term from Eq.~(\ref{logderivative_cdd}) to the
simple virial estimator (\ref{qr_est_virial}) becomes
\begin{eqnarray}
\frac{\partial V_{\text{constr}}}{\partial m}&=&
\frac{2P}{\hbar^2\beta^2}\frac{|d}{\D\Delta m}\,(m+\Delta m)
\left[\frac{\xi\left(r_{\text{resc}}\right)}
{\partial_{r_{\text{resc}}}\xi(r_{\text{resc}})}\right]^2\,,\label{cdd_est_virial}\\
r_{\text{resc}}&\equiv&\left(\frac{m}{m+\Delta
m}\right)^{1/2}r\,.\nonumber
\end{eqnarray}
Finally, if we rescale coordinates according to Eq.
(\ref{rescaling_subtract}), or better, as
$$
x^{(s)}\equiv
m^{1/2}(r^{(s)}-r^{c})\quad\text{for~}\;s=1,\,\ldots,\,P\,,
$$
we obtain the same estimator as (\ref{cdd_est_virial}), only the
rescaled coordinate is defined as
\begin{equation}
r_{\text{resc}}^{(s)}\equiv r^c+\left(\frac{m}{m+\Delta
m}\right)^{1/2}(r^{(s)}-r^c)\,.
\label{cdd_est_virial_generalized}
\end{equation}

The generalization to more-dimensional systems is fairly
straightforward. Only in the case of the simple virial estimator
(\ref{qr_est_virial}) or (\ref{cdd_est_virial}), care must be
taken to account for the unbound (free) degrees of freedom by
appropriately rescaling the corresponding volume. For instance,
for bimolecular reactions, the potential in the reactant region is
independent of the center-of-mass-coordinate and the relative
coordinate of the two molecules. For details, see
Ref.~\cite{vanicek:2005b}.

\section{Numerical results}

In Ref.~\cite{vanicek:2005}, the QI procedure for evaluating KIEs
was successfully tested on several problems of increasing
complexity: the one-dimensional Eckart barrier and the isotopic
variants of both the collinear and the three-dimensional
hydrogen-exchange reaction H+H$_2\rightarrow$ H$_2$+H. The results
for the KIE = $k$(H+H$_2$)/$k$(D+D$_2$) as a function of the
inverse temperature $1/T$ for both collinear and three-dimensional
versions of the reaction are also shown here in Fig.~1. The figure
compares the exact quantum-mechanical result
\cite{vanicek:2005,truhlar:1973} with the results of the QI, SQI,
and TST approximations. The three-dimensional version also shows
the result of the canonical variational TST with semiclassical
tunneling correction (CVT) \cite{fast:1998}. In general, the
results of the QI approximation are very good: the error is
smaller than $10\%$ for temperatures 250 to 600~K. For lower
temperatures, the larger error is due to using a single dividing
surface: results can be improved by considering two separate
dividing surfaces. At high temperatures, the error is due to
classical recrossing and cannot be corrected within the QI
approximation. For further details of the calculation see
Ref.~\cite{vanicek:2005}.

\begin{figure}
\centerline{\resizebox{\hsize}{!}{\includegraphics{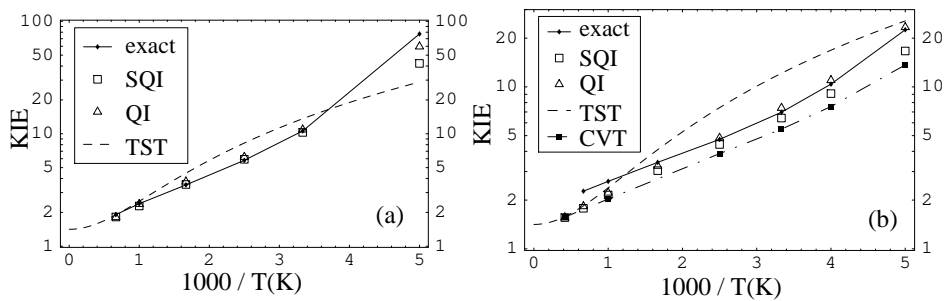}}}
\vskip-5mm
\caption{Kinetic isotope effect $k$(H+H$_2$)/$k$(D+D$_2$) for the
hydrogen exchange reaction: (a) the collinear version, (b) the
three-dimensional version.}
\end{figure}

Three types of estimators for the logarithmic derivative of $Q_r$,
given in Eqs. (\ref{qr_est_thermodynamic}), (\ref{qr_est_virial}),
and (\ref{qr_est_virial_centroid}), are compared in Fig.~2. This
calculation is for the collinear version of the KIE =
$k$(H+H$_2$)/$k$(D+D$_2$) at 300 K. The calculation was done with
10 walkers and a fixed number $10^5$ Monte-Carlo moves for all
$P$. The left part of the figure shows the convergence of the
partition-function ratio $Q_{r}(1)/Q_{r}(0)$ as a function of the
number of slices $P$. The ratio was computed via the thermodynamic
integration (\ref{charging_Qr}) in which the three different
estimators (\ref{qr_est_thermodynamic}), (\ref{qr_est_virial}),
and (\ref{qr_est_virial_centroid}) for $\dfrac{\D\log
Q_r}{\D\lambda}$ were used. The right panel shows the
$P$-dependence of the relative error of $Q_{r}(1)/Q_{r}(0)$. As
expected, for large $P$, the error of the generalized virial
estimator (\ref{qr_est_virial_centroid}) is almost independent of
$P$.  On the other hand the error of the thermodynamic estimator
(\ref{qr_est_thermodynamic}) grows with $P$. Even for small $P$,
the generalized virial estimator is superior. Finally, we can see
the importance of subtracting the centroid motion before rescaling
in Eq. (\ref{rescaling_subtract}) by comparing errors of the
simple (\ref{qr_est_virial}) and generalized
(\ref{qr_est_virial_centroid}) virial estimators. The difference
is due to the fact that we have two free degrees of freedom in the
reactant region of the collinear bimolecular reaction. Similar
conclusions (not shown here) can be obtained for the ratio
$C_{\mathrm{dd}}(1)/C_{\mathrm{dd}}(0)$, except that the error of
the generalized virial estimator has a weak dependence on $P$
arising from the additional term due to the constraint to the
dividing surfaces.

\begin{figure}[t]
\centerline{\resizebox{\hsize}{!}{\includegraphics{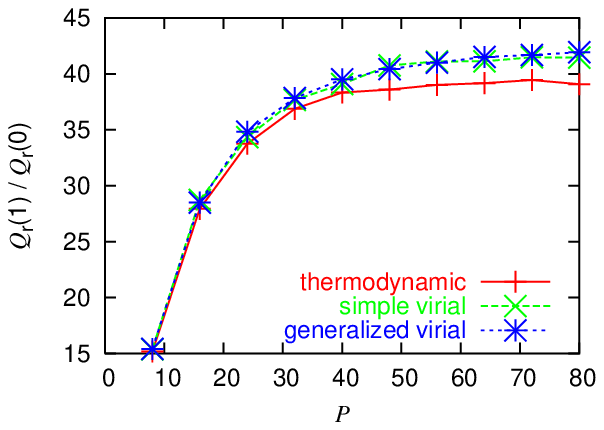}\includegraphics{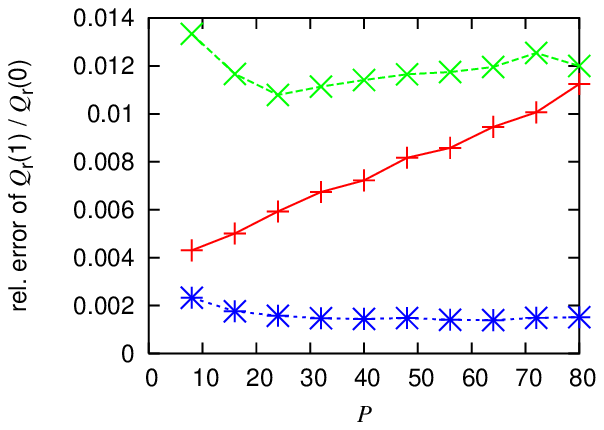}}}
\vskip-5mm
\caption{Comparison of the thermodynamic, virial, and generalized
virial estimators for the logarithmic derivative of $Q_r$. Left:
ratio of the partition functions as a function of $P$, right: its
relative error as a function of $P$.}
\end{figure}

\section{Conclusion}

Judging from the numerical results in the previous section, the QI
approach for computing kinetic isotope effects is very promising.
The procedure is quite general: multi-dimensional estimators for
all relevant quantities are presented in Refs.
\cite{vanicek:2005,vanicek:2005b}. Although the path-integral
approach has been chosen because of its favorable scaling with the
number of degrees of freedom, the computations for
many-dimensional systems are still difficult.

One obstacle is the difficulty of efficient sampling of a
many-dimensional configuration space. With proper estimators, we
may decrease statistical and systematic discretization errors, but
it is difficult to avoid systematic errors due to long
correlations. For this reason, it may be efficient to use a
different number of imaginary time slices \cite{li:2005} for
different degrees of freedom, which is a generalization of more
crude mixed quantum-classical methods.

Another obstacle to obtaining a good match between theory and
experiment is the potential energy surface for the reaction. While
more accurate \textit{ab initio} potentials are computationally
very expensive, the much faster molecular-mechanics force fields
are often too crude. In Ref.~\cite{vanicek:2005b}, in which the QI
method is used to compute the KIE for the isomerization of
\textit{cis}-pentadiene, two approaches are taken: in one, an
empirical valence bond (EVB) potential is formed from the
equilibrium potentials for reactants and products; in the other, a
more accurate but also a more-expensive semi-empirical potential
is used. Because of the computational expense already for this
system with 39 degrees of freedom, it appear that due to their
better accuracy such semi-empirical potentials will be the
potentials of choice for intermediate-size systems, and the EVB
potentials based on molecular-mechanical force fields the
potentials of choice for truly many-dimensional systems.

\vfill\eject

\noindent{\small J. Van\'{\i}\v{c}ek would like to thank Y. Li and
C. Predescu for useful discussions. This work was supported by the
Office of Naval Research Grant No. N00014-05-1-0457 and by the
Director, Office of Science, Office of Basic Energy Sciences,
Chemical Sciences, Geosciences, and Biosciences Division, U.S.
Department of Energy under Contract No. DE-AC02-05CH11231.}


\begin{thebibliography}{99}
\bibitem{kohen:1999}
A. Kohen, R. Cannio, S. Bartolucci and J.P. Klinman: Nature
\textbf{399} (1999) 496.
\bibitem{melander:1960}
L.C.S. Melander: \textit{Isotope Effects on Reaction Rates}.
Ronald Press, New York, 1960.
\bibitem{collins:1971}
C.J. Collins and N.S. Bowman: Eds.: \textit{Isotope effects in
chemical reactions}. Van Nostrand Reinhold, New York, 1971.
\bibitem{miller:1975}
W.H. Miller: J. Chem. Phys. \textbf{62} (1975) 1899.
\bibitem{marcus:1977}
R.A. Marcus and M.E. Coltrin: J. Chem. Phys. \textbf{67} (1977)
2609.
\bibitem{truhlar:1984}
D.J. Truhlar and B.C. Garrett: Annu. Rev. Phys. Chem. \textbf{35}
(1984) 159.
\bibitem{miller:1998}
W.H. Miller: Faraday Discuss. \textbf{110} (1998) 1.
\bibitem{voth:1989}
G.A. Voth, D. Chandler and W.H. Miller: J. Chem. Phys. \textbf{91}
(1989) 7749.
\bibitem{hansen:1996}
N.F. Hansen and H.C. Andersen: J. Phys. Chem. \textbf{100} (1996)
1137.
\bibitem{truong:1993}
T.N. Truong, D. Lu, G.C. Lynch \textit{et al.}: Comput. Phys.
Commun. \textbf{75} (1993) 143.
\bibitem{miller:2003}
W.H. Miller, Y. Zhao, M. Ceotto and S. Yang: J. Chem. Phys.
\textbf{119} (2003) 1329.
\bibitem{vanicek:2005}
J. Van\'{\i}\v{c}ek, W.H. Miller, J.F. Castillo and F.J. Aoiz: J.
Chem. Phys. \textbf{123} (2005) 054108.
\bibitem{vanicek:2005b}
J. Van\'{\i}\v{c}ek and W.H. Miller: in preparation.
\bibitem{ceotto:2003}
M. Ceotto and W.H. Miller: private communication.
\bibitem{miller:1983}
W.H. Miller, S.D. Schwartz and J.W. Tromp: J. Chem. Phys.
\textbf{79} (1983) 4889.
\bibitem{yamamoto:2004}
T. Yamamoto and W.H. Miller: J. Chem. Phys. \textbf{120} (2004)
3086.
\bibitem{zhao:2004}
Y. Zhao, T. Yamamoto and W.H. Miller: J. Chem. Phys. \textbf{120}
(2004) 3100.
\bibitem{chandler:1987}
D. Chandler: \textit{Introduction to Modern Statistical
Mechanics}. Oxford University Press, New York, 1987.
\bibitem{barker:1979}
J. Barker: J. Chem. Phys. \textbf{70} (1979) 2914.
\bibitem{herman:1982}
M.F. Herman, E.J. Bruskin and B.J. Berne: J. Chem. Phys.
\textbf{76} (1982) 5150.
\bibitem{parrinello:1984}
M. Parrinello and A. Rahman: J. Chem. Phys. \textbf{80} (1984)
861.
\bibitem{predescu:2003}
C. Predescu, D. Sabo, J.D. Doll and D.L. Freeman: J. Chem. Phys.
\textbf{119} (2003) 12119.
\bibitem{predescu:2004}
C. Predescu: Phys. Rev. E \textbf{70} (2004) 066705.
\bibitem{truhlar:1973}
D.G. Truhlar, A. Kuppermann and J.T. Adams: J. Chem. Phys.
\textbf{59} (1973) 395.
\bibitem{fast:1998}
P.L. Fast, J.C. Corchado and D.G. Truhlar: J. Chem. Phys.
\textbf{109} (1998) 6237.
\bibitem{li:2005}
Y. Li and W.H. Miller: Mol. Phys. \textbf{103} (2005) 203.
\end{thebibliography}
\end {document}